\baselineskip 16pt
\font\carbig=cmr10 scaled 2074

\magnification=\magstep0
\hsize=15.0truecm
\hoffset=0.6truecm
\vsize=22.0truecm
\hyphenation{IL-LU-MI-NA-TI}

\def\sqr#1#2{{\vcenter{\vbox{\hrule height.#2pt
    \hbox{\vrule width.#2pt height#1pt \kern#1pt
    \vrule width.#2pt}\hrule height.#2pt}}}}

\def\boxit#1{\vbox{\hrule\hbox{\vrule\kern5pt
     \vbox{\kern5pt#1\kern3pt}\kern5pt\vrule}\hrule}}

\def\E{{\bf E}}
\def\e{{\rm e}}

\def\RE{{\bf R}}

\def\OM{{\Omega}}
\def\F{{\cal F}}
\def\P{{\bf P}}

\pageno=1
\item{}
\vskip 1cm
\centerline{\carbig Controlled quantum evolutions}
\vskip 5pt
\centerline{{\carbig and stochastic mechanics}%
         \footnote{}
{\baselineskip 12pt Paper presented at the 7$^{th}$ UK Conference on
{\it Mathematical and Conceptual Foundations of Modern Physics}; 
Nottingham (UK) 7-11 
September, 1998.}}
\vskip 12pt
\centerline{\bf Nicola Cufaro Petroni}
\par
\centerline{{\it INFN Sezione di Bari and Dipartimento di Fisica 
                 dell'Universit\`a di Bari,}}
\par
\centerline{{\it via Amendola 173, 70126 Bari (Italy)}}
\par
\centerline{CUFARO@BARI.INFN.IT}
\vskip 7pt
\centerline{{\bf Salvatore De Martino, Silvio De Siena and 
Fabrizio Illuminati}}
\par
\centerline{{\it INFN Sezione di Napoli - Gruppo collegato di Salerno and}}
\par
\centerline{{\it Dipartimento di Fisica dell'Universit\`a di Salerno}}
\par
\centerline{{\it via S.Allende, 84081 Baronissi, Salerno (Italy)}}
\par
\centerline{DEMARTINO@PHYSICS.UNISA.IT, DESIENA@PHYSICS.UNISA.IT}
\par
\centerline{ILLUMINATI@PHYSICS.UNISA.IT}
\vskip 1.3cm\noindent
{\bf ABSTRACT: {\it We perform a detailed analysis of the non stationary 
solutions 
of the evolution (Fokker-Planck) equations associated to either stationary or 
non stationary quantum states by the stochastic mechanics. For the excited 
stationary states of quantum systems with singular velocity fields we 
explicitely discuss the exact solutions for the HO case. 
Moreover the possibility of modifying the original potentials in order to 
implement arbitrary evolutions ruled by these equations 
is discussed with respect 
to both possible models for quantum measurements and 
applications to the control 
of particle beams in accelerators.}}
\vskip 1.3cm
\centerline{\bf 1. Introduction}
\vskip 10pt\noindent
In a few papers [1] the analogy between diffusive classical systems and
quantum systems has been reconsidered from the standpoint of the stochastic
mechanics (SM) [2], [3], and particular attention was devoted 
there to the evolution 
of the classical systems associated to a quantum wave function
when the conditions imposed by the stochastic variational principle are not
satisfied (non extremal processes). The hypothesis that the 
evolving distribution 
converges in time toward the quantum distribution, 
constituted several years ago an 
important point in the answer by Bohm and Vigier 
to some criticisms to the assumptions of the Causal Interpretation of the 
Quantum Mechanics (CIQM) [4]. In the quoted papers [1] it was pointed 
out that, while the right convergence was in fact achieved for a few quantum
examples, these results could not be considered general as shown in some  
counterexamples: in fact not only for particular non stationary wave functions 
(as for a minimal uncertainty packet), but also for stationary 
states with nodes 
(namely with zeros) we do not seem to get the right asymptotic behaviour. 
For stationary states with nodes the problem is that the corresponding velocity 
field to consider in the Fokker-Planck equation shows singularities 
in the locations 
of the nodes of the wave function. These singularities effectively 
separate the available 
interval of the space variables into (probabilistically) non 
communicating sections 
which trap any amount of probability initially 
attributed and make the system non ergodic.  
\par
In a more recent paper [5] it has been shown first of all
that for transitive systems with stationary velocity 
fields (as, for example, a stationary state without nodes) we always have an
exponential convergence to the right quantum probability distribution associated
to the extremal process, even if we initially start from an 
arbitrary non extremal process. 
These results can also be extended to an arbitrary stationary state if we
separately consider the process as confined in every configuration space
region between two subsequent nodes. 
Moreover it has been remarked there 
that while the non extremal processes should be considered virtual, 
as trajectories in
the classical Lagrangian mechanics, they can also be turned real 
if we modify the potential in a suitable way. The interest of this remark
lies not only
in the fact that non extremal processes are exactly what is lacking 
in quantum mechanics
in order to interpret it as a 
totally classical stochastic process theory (for example
in order to have a classical picture of a double slit experiment [6]),
but also in the possibility of engineering
some new controlled real evolutions of quantum states.
In fact this could be useful to study (a) transitions between stationary states 
(b) possible models for measure theory [3] and (c) control of
the particle beam dynamics in accelerators [7].
In a sense the SM 
is also a theory, independent from quantum mechanics, with
applications in several physical fields, in particular for systems not perfectly
described by the quantum formalism, but whose evolution is correctly 
controlled by 
quantum fluctuation: the so called mesoscopic or quantum-like systems. 
This behaviour 
characterizes, for example, the beam dynamics in particle accelerators and
there is evidence that it could be
described by the stochastic formalism of Nelson diffusions [1], [7].
Of course in this model trajectories and transition probabilities always are 
perfectly meaningful and,
to study in detail the evolution of the probability 
distributions, and in particular to try to understand if and how 
it is possible to realize controlled evolutions, 
it is necessary to determine the fundamental solutions (transition 
probability densities) associated by SM to every 
quantum state in consideration: a problem dealt with in the following sections.
\vskip 30pt
\centerline{\bf 2. Fokker-Planck equations for stochastic mechanics}
\vskip 10pt\noindent
SM is a generalization of classical mechanics based
on the theory of classical stochastic processes [2]. The variational 
principles of Lagrangian type provide a foundation for it, as for the classical 
mechanics or the field theory [3]. In this scheme
the deterministic trajectories of classical mechanics
are replaced by the random trajectories of diffusion processes in the
configuration space. The surprisig feature is that programming equations
derived from the stochastic version of the lagrangian principle are formally 
identical to the equations of a Madelung fluid [8], the hydrodynamical 
equivalent of the Schr\"odinger equation in the Stochastic Interpretation of 
the Quantum Mechanics (SIQM) [9]. On this basis, it is possible to 
develop an interpretative scheme where the phenomenological predictions of 
SM coincide with that of quantum mechanics for all the 
experimentally measurable quantities. Within this interpretative code the
SM is nothing but a quantization procedure, different from 
the ordinary ones only formally, but completely equivalent from the point of 
view of the physical consequences. Hence 
we consider here the SM
as a probabilistic simulation of quantum mechanics, providing a bridge between 
this fundamental section of physics and the stochastic differential calculus.
However it is well known that 
the most peculiar features of the involved stochastic processes, namely
the transition probability densities, seem not always enter into this code 
scheme: in fact, if we want to check experimentally if the transition 
probabilities are the right ones for a given quantum state, 
we are obliged to perform 
repeated position measurements on the quantum system; but, according to 
quantum theory, the quantum state changes at every measurement (wave packet
reduction), and since our transition probabilities are associated to a 
well defined wave function it will be in general practically impossible 
to experimentally observe a well defined transition probability.
Several ways out of these difficulties have been explored: 
for example stochastic 
mechanic scheme could be modified by means of non constant diffusion 
coefficients [1]; or alternatively it would be possible to modify the 
stochastic evolution during the measurement [10]. 
Here we will rather assume that the 
processes which do not satisfy the stochastic variational principle 
still keep a physical meaning and that tey
will rapidly converge (in time) toward the 
processes associated to quantum states. Indeed
on the one hand any departure from the 
distributions of quantum mechanics will quickly be reabsorbed in the time 
evolution, at least in many meaningful cases; and on the other hand 
the non standard evolving distributions could be realized by 
suitable quantum systems for modified, time dependent potentials which may
asymptotically in time rejoin the usual potentials.
\par
SM is a model intended to achieve a
connection between quantum mechanics and classical random
phenomena: here we will recall a few notions in order to fix the notation.
The position of a classical particle is promoted to a vector Markov 
process
$\xi(t)$ defined on some probabilistic space $(\OM,\F,\P)$ and taking values
in $\RE^3$. We suppose that this process is characterized by a
pdf $f({\bf r},t)$ and a transition pdf $p({\bf r},t|\,{\bf r'},t')$ and
satisfies an It\^o stochastic differential equation of the form
$$d\xi_j(t)=v_j\bigl(\xi(t),t\bigr)dt+d\eta_j(t) \eqno(2.1)$$
where $v_j$ are the components of
the forward velocity field. However here $v_j$ are 
not given a priori, but play the role of dynamical variables and are
subsequently determined on the basis of a variational principle, namely
on the basis of a dynamics. On the other hand $\eta(t)$ is a 
Brownian process independent of $\xi(t)$ and such that
$$\E_t\bigl(d\eta_j(t)\bigr)=0\,,\qquad
  \E_t\bigl(d\eta_j(t)\,d\eta_k(t)\bigr)=2\,D\, \delta_{jk}\,dt\eqno{2.2}$$
where $d\eta(t)=\eta(t+dt)-\eta(t)$ (for $dt>0$), $D$ is a diffusion 
coefficient, and $\E_t$ are the
conditional expectations with respect to $\xi(t)$.
In what follows we will limit ourselves to the case of the one dimensional
trajectories, so that the Markov processes $\xi(t)$ considered will
always take values in $\RE$. Moreover we will suppose for the time being
that the forces acting on the particle will be defined by means of 
a time-independent potential $V(x)$.
A suitable definition of the Lagrangian and of the stochastic action 
functional for the system described by the dynamical variables
$f$ and $v$ allows us to select, by means of the principle of
stationarity of the action, the processes which reproduce the 
quantum mechanics [2], [3]. 
In fact, while the pdf $f(x,t)$ of the process satisfies, as usual, the
Forward Fokker-Planck (FP) equation associated to (2.1)
$$\partial_tf=D\partial_x^2f-\partial_x(vf)=\partial_x(D\partial_xf-vf)\,,
                                        \eqno{(2.3)}$$
the following choice for the Lagrangian field
$${\cal L}(x,t)={m\over2}v^2(x,t)+mD\partial_xv(x,t)-V(x)\eqno{(2.4)}$$
enables us to define a stochastic action funcional
$${\cal A}=\int_{t_0}^{t_1}\E{\cal L}\big(\xi(t),t\big)\,dt\eqno{(2.5)}$$
which leads, through the stationarity condition 
$\delta {\cal A}=0$, to the equation
$$\partial_tS+{(\partial_xS)^2\over2m}+V\,-2mD^2\,
                    {\partial^2_x\sqrt{f}\over \sqrt{f}}=0\eqno{(2.6)}$$
involving a field $S(x,t)$ defined as
$$S(x,t)=-\int_t^{t_1}
         \E\left({\cal L}\big(\xi(s),s\big)\,\big|\,\xi(t)=x\right)\,ds
        +\E\left(S_1\big(\xi(t_1)\big)\,\big|\,\xi(t)=x\right) \eqno{(2.7)}$$
where $S_1(\,\cdot\,)=S(\,\cdot\,,t_1)$ is an arbitrary final condition.
Now the relevant remark is that if $R(x,t)=\sqrt{f(x,t)}$, and if we define 
$$\psi(x,t)=R(x,t)\,{\rm e}^{iS(x,t)/\hbar}\eqno{(2.8)}$$
the equation (2.6) takes the form
$$\partial_tS+{(\partial_xS)^2\over2m}+V\,-\,{\hbar^2\over2m}\,
                    {\partial^2_xR\over R}=0\,,\eqno{(2.9)}$$
and the complex wave function $\psi$ will satisfy the Schr\"odinger equation
$$i\hbar\partial_t\psi=\hat H\psi=-\,{\hbar^2\over 2m}\,\partial_x^2\psi
                            +V\psi\,,\eqno{(2.10)}$$
provided that the diffusione coefficient be connected to the Planck constant
by the relation
$$D={\hbar\over 2m}\,.\eqno{(2.11)}$$
\par
This trail leading from classical stochastic processes (plus a dynamics) to
quantum mechanics can also be trod in the reverse way following the line
of reasoning of the SIQM which, as
it is well known, is formally ruled by the same differential equations
as the SM. If we start from 
the (one dimensional) Schr\"odinger equation (2.10) 
with the {\it Ansatz} (2.8), and if we separate the real and the
imaginary parts as usual in SIQM [8],
the function $f=R^2=|\psi|^2$ comes out
to be a particular solution of a FP equation  
of the form (2.3) with constant diffusion coefficient (2.11) and
forward velocity field 
$$v(x,t)={1\over m}\,\partial_xS+{\hbar\over 2m}\,\partial_x(\ln R^2)\,.
                                          \eqno{(2.12)}$$
On the other hand the explicit dependence of $v$ on  the form of $R$ clearly
indicates that to have a solution of (2.3) which makes quantum sense we
must pick-up just one, suitable, particular solution. 
In fact the system is ruled not 
only by the
FP equation (2.3), but also by the second, dynamical equation
(2.9), the so-called Hamilton-Jacobi-Madelung (HJM) equation,
deduced by separating the real and imaginary parts of (2.10) (see [8]). 
The analogy between (2.3) and a FP equation, which looks
rather accidental in a purely SIQM context, is more than formal 
since, as we have
briefly recalled, the SM shows how to recover 
both the equations (2.3) and (2.9) (and hence the Schr\"odinger equation (2.10))
in a purely classical, dynamical stochastic context.
\vskip 30pt
\centerline{\bf 3. The eigenvalue problem for the FP equation}
\vskip 10pt\noindent
Let us recall here (see for example [11])
a few generalities about the pdf's 
(probability density functions) $f(x,t)$ solutions of a 
one-dimensional FP equation
of the form
$$\partial_tf=\partial_x^2(Df)-\partial_x(vf)=
     \partial_x\bigl[\partial_x(Df)-vf\bigr]\eqno{(3.1)}$$
defined for $x\in[a,b]$ and $t\geq t_0$, when $D(x)$ and $v(x)$ are two time 
independent functions such that $D(x)>0$, $v(x)$ has no singularities in 
$(a,b)$, and both are continuous and differentiable functions.
The conditions imposed on the probabilistic solutions are of course
$$\eqalign{f(x,t)\geq0\,,&\qquad\qquad a<x<b\,,\;t_0\leq t\,,\cr
     \int_a^bf(x,t)\,dx=1\,,&\qquad\qquad t_0\leq t\,,\cr}\eqno{(3.2)}$$
and from the form of (3.1) the second condition also takes the form
$$\bigl[\partial_x(Df)-vf\bigr]_{a,b}=0\,,\qquad t_0\leq t\,.\eqno{(3.3)}$$
Suitable initial conditions will be added to produce the required 
evolution: for example the transition pdf $p(x,t|x_0,t_0)$ will be selected by 
the initial condition
$$\lim_{t\to t_0^+}f(x,t)=f(x,t_0^+)=\delta(x-x_0)\,.\eqno{(3.4)}$$
It is also possible to show by direct calculation that
$$h(x)=N^{-1}\,\e^{-\int[D'(x)-v(x)]/D(x)\,dx}\,,\qquad
       N=\int_a^b\e^{-\int[D'(x)-v(x)]/D(x)\,dx}\,dx\eqno{(3.5)}$$
is an invariant (time independent) solution of (3.1) satisfying the conditions
(3.2). Remark however that (3.1) 
is not in the standard self-adjoint form [12];
but if we define the function $g(x,t)$ by means of
$$f(x,t)=\sqrt{h(x)}\,g(x,t)\eqno{(3.6)}$$
it would be easy to show that $g(x,t)$ obeys now an equation of 
the form
$$\partial_tg={\cal L}g\eqno{(3.7)}$$
where the operator ${\cal L}$ defined by
$${\cal L}\varphi=
     {d\over dx}\left[p(x)\,{d\varphi(x)\over dx}\right]-q(x)\varphi(x)\,,
               \eqno{(3.8)}$$
with
$$\eqalign{p(x)&=D(x)>0\,,\cr
           q(x)&={\bigl[D'(x)-v(x)\bigr]^2\over4D(x)}\,-\,
                   {\bigl[D'(x)-v(x)\bigr]'\over2}\,,\cr}\eqno{(3.9)}$$
is now self-adjoint. 
Then, by separating the variables by means of
$g(x,t)=\gamma(t)G(x)$
we have $\gamma(t)={\rm e}^{-\lambda t}$ while $G$ must 
be solution of a typical Sturm-Liouville problem associated to 
the equation
$${\cal L}G(x)+\lambda G(x)=0\eqno{(3.10)}$$
with the boundary conditions 
$$\eqalign{&\bigl[D'(a)-v(a)\bigr]G(a)+2D(a)G'(a)=0\,,\cr
           &\bigl[D'(b)-v(b)\bigr]G(b)+2D(b)G'(b)=0\,.\cr}\eqno{(3.11)}$$
It easy to see that $\lambda=0$ is always an eigenvalue for the problem (3.10) 
with (3.11), and that the corresponding eigenfunction is $\sqrt{h(x)}$ 
as defined from (3.5).
\par
For the differential problem (3.10)
with (3.11) we have that [12] the simple eigenvalues $\lambda_n$ 
will constitute an infinite, increasing sequence and the corresponding 
eigenfunction $G_n(x)$ will have $n$ simple zeros in $(a,b)$. For us this means 
that $\lambda_0=0$, corresponding to the eigenfunction $G_0(x)=\sqrt{h(x)}$ 
which never vanishes in $(a,b)$, is the lowest eigenvalue and that all other 
eigenvalues are strictly positive. Moreover the eigenfunctions will 
constitute a complete orthonormal set of functions in $L^2\bigl([a,b]\bigr)$
[13]. As a consequence the general solution of (3.1) with (3.2) 
will have the form
$$f(x,t)=\sum_{n=0}^{\infty}c_n\e^{-\lambda_nt}\sqrt{h(x)}G_n(x)\eqno{(3.12)}$$
with $c_0=1$ for normalization (remember that $\lambda_0=0$). The coefficients 
$c_n$ for a particular solution are selected by an initial condition 
$$f(x,t_0^+)=f_0(x)\eqno{(3.13)}$$
and are then calculated from the orthonormality relations as
$$c_n=\int_a^bf_0(x)\,{G_n(x)\over\sqrt{h(x)}}\,dx\,.\eqno{(3.14)}$$
In particular for the transition pdf we have from (3.4) that
$$c_n={G_n(x_0)\over\sqrt{h(x_0)}}\,.\eqno{(3.15)}$$
\par\noindent
Since 
$\lambda_0=0$ and $\lambda_n>0$ for $n\geq1$, the general 
solution (3.12) of (3.1) has a precise time evolution: all the 
exponential factors in (3.12) vanish with $t\to+\infty$ with the only 
exception of the term $n=0$ which is constant, so that exponentially fast we 
will always have
$$\lim_{t\to+\infty}f(x,t)=c_0\sqrt{h(x)}G_0(x)=h(x)\,,\eqno{(3.16)}$$
namely the general solution will always relax in time toward the invariant 
solution $h(x)$.
\vskip 30pt
\centerline{\bf 4. Stationary quantum states}
\vskip 10pt\noindent
Let us consider now a Schr\"odinger equation (2.10) with 
a time-independent potential $V(x)$ which gives rise to a purely
discrete spectrum and bound, normalizable states, and let us use the
following notations for stationary states, eigenvalues and eigenfunctions:
$$\eqalign{\psi_n(x,t)&=\phi_n(x)\,{\rm e}^{-iE_nt/\hbar}\cr
           \hat H\phi_n&=-{\hbar^2\over 2m}\,\phi''_n+V\phi_n=E_n\phi_n\,.\cr}
                                                    \eqno{(4.1)}$$
Taking into account the relation (2.11) the previous
eigenvalue equation can also be recast in the following form
$$D\phi''_n={V-E_n\over\hbar}\,\phi_n\,.\eqno{(4.2)}$$
For these stationary states the pdf is
the time independent, real function
$$f_n(x)=|\psi_n(x,t)|^2=\phi_n^2(x)\,,\eqno{(4.3)}$$
and 
$$S(x,t)=-E_nt\,,\qquad R(x,t)=\phi_n(x)\,,\eqno{(4.4)}$$
so that for our state the velocity field is
$$v_n(x)=2D\,{\phi'_n(x)\over\phi_n(x)}\,.\eqno{(4.5)}$$
This means that now $v_n$ is time-independent and it presents
singularities in the zeros (nodes) of the eigenfunction. Since the $n$-th
eigenfunction of a quantum system with bound states has exactly $n$ simple
nodes [12]
that we will indicate with $x_1,\dots,x_n$, the coefficients of the
FP equation (2.3) are not defined in these $n$ points and we
will be obliged to solve it in separate intervals by imposing the
right boundary conditions connecting the different sections. In fact these
singularities effectively separate the real axis in $n+1$ 
sub-intervals with walls impenetrable to the probability current.
Hence the process will not have an unique invariant measure and will never 
cross the boundaries fixed by the singularities of $v(x)$: if we start in one 
of the intervals in which the axis is so divided we will always remain there
[14].
\par
As a consequence we must think the normalization
integral (3.2) (with $a=-\infty$ and $b=+\infty$) 
as the sum of $n+1$ integrals over the sub-intervals $[x_k,x_{k+1}]$
with $k=0,1,\dots,n$ (where we understand, to unificate the notation, that
$x_0=-\infty$ and $x_{n+1}=+\infty$). Hence for $n\geq1$ we will be obliged
to solve the equation (2.3) in every interval $[x_k,x_{k+1}]$ by requiring
that the integrals
$$\int_{x_k}^{x_{k+1}}f(x,t)\,dx\eqno{(4.6)}$$
be kept at a constant value for $t\geq t_0$: this value is not, in general,
equal to one (only the sum of these $n+1$ integrals amounts to one) and,
since the separate intervals can not communicate, it will be fixed by the
choice of the initial conditions. 
Hence the boundary conditions associated to (2.3) require 
the conservation of 
the probability in $[x_k,x_{k+1}]$, namely the vanishing of the 
probability current at the end points of the interval:
$$\bigl[D\partial_xf-vf\bigr]_{x_k,x_{k+1}}=0\,,\qquad t\geq t_0\,.
                                    \eqno{(4.7)}$$
To have a particular solution we must moreover specify
the initial conditions: in particular we will be interested in the 
transition pdf $p(x,t|x_0,t_0)$, 
which is singled out by the initial condition (3.4),
since [1] the asymptotic approximation in $L^1$ among
solutions of (2.3) is ruled by the asymptotic behavior of $p(x,t|x_0,t_0)$
through the Chapman-Kolmogorov equation
$$f(x,t)=\int_{-\infty}^{+\infty}p(x,t|y,t_0)f(y,t_0^+)\,dy\,.\eqno{(4.8)}$$
It is clear at this point that in every interval $[x_k,x_{k+1}]$ (both finite
or infinite) we
can solve the equation (2.3) along the guidelines sketched in the section 3
by keeping in mind that in $[x_k,x_{k+1}]$ we already know the invariant,
time-independent solution $\phi_n^2(x)$ (or, more precisely, its restriction
to the said interval) which is never zero in this interval
with the exception of the extremes $x_k$ and $x_{k+1}$. Hence, as we have
seen in the general case, with the position
$$f(x,t)=\phi_n(x)g(x,t)\eqno{(4.9)}$$
we can reduce (2.3) to the form
$$\partial_tg={\cal L}_ng\eqno{(4.10)}$$
where ${\cal L}_n$ is now the 
self-adjoint operator defined on $[x_k,x_{k+1}]$ by
$${\cal L}_n\varphi(x)={{\rm d}\over{\rm d}x}
        \left[p(x){{\rm d}\varphi(x)\over{\rm d}x}\right]
                     -q_n(x)\varphi(x)\eqno{(4.11)}$$
where we have now
$$p(x)=D>0\,;\qquad q_n(x)={v_n^2(x)\over 4D}+{v_n'(x)\over 2}\,.\eqno{(4.12)}$$
To solve (4.10) it is in general advisable to separate the variables,
so that we immediately have $\gamma(t)={\rm e}^{-\lambda t}$ while $G$ must 
be solution of the Sturm-Liouville problem associated to 
the equation
$${\cal L}_nG(x)+\lambda G(x)=0\eqno{(4.13)}$$
with the boundary conditions 
$$\bigl[2DG'(x)-v_n(x)G(x)\bigr]_{x_k,x_{k+1}}=0\,.\eqno{(4.14)}$$
The general behaviour of the solutions obtained as expansions in the system of 
the eigenfunctions of (4.13) has already been discussed in section 3.
In particular we deduce from (3.12) that for the stationary quantum 
states (more precisely, in every subinterval defined by two subsequent nodes)
all the solutions of (2.3) always converge in time toward the right quantum
solution $|\phi_n|^2$: a general result not contained in the previous 
papers [1]. As a further consequence a quantum solution $\phi^2_n$ defined
on the entire interval $(-\infty,+\infty)$ will be stable under deviations
from its initial condition.
\vskip 30pt
\centerline{\bf 5. Harmonic oscillator}
\vskip 10pt\noindent
To see in an explicit way how the pdf's of SM evolve, let us
consider now in detail the particular example of a quantum harmonic 
oscillator (HO) characterized by the potential
$$V(x)={m\over2}\,\omega^2x^2\,.\eqno{(5.1)}$$
It is well-known that its eigenvalues are
$$E_n=\hbar\omega\left(n+{1\over2}\right)\,;\qquad n=0,1,2\dots\eqno{(5.2)}$$
while, with the notation
$$\sigma_0^2={\hbar\over2m\omega}\,,\eqno{(5.3)}$$
the eigenfuncions are
$$\phi_n(x)=
    {1\over\sqrt{\sigma_0\sqrt{2\pi}2^nn!}}\,{\rm e}^{-x^2/4\sigma_0^2}\,
                 H_n\left({x\over\sigma_0\sqrt{2}}\right)\eqno{(5.4)}$$
where $H_n$ are the Hermite polynomials.
The corresponding velocity fields are easily calculated and are for example
$$\eqalign{v_0(x)&=-\omega x\,,\cr
         v_1(x)&=2\,{\omega\sigma_0^2\over x}-\omega x\,,\cr
         v_2(x)&=4\omega\sigma_0^2\,{x\over x^2-\sigma_0^2}-\omega x\,,\cr}
                                            \eqno{(5.6)}$$
with singularities in the zeros $x_k$ of the Hermite polynomials. 
If we now keep the form of the velocity fields fixed we can consider (2.3) as
an ordinary FP equation for a diffusion process
and solve it to see the approach to the 
equilibrium of the general solutions.
When $n=0$ the equation (2.3) takes the form
$$\partial_tf=\omega\sigma_0^2\partial_x^2f+\omega x\partial_xf+\omega f
                 \eqno{(5.7)}$$
and the fundamental solution comes out to be the Ornstein-Uhlenbeck transition
pdf
$$p(x,t|x_0,t_0)={1\over\sigma(t)\sqrt{2\pi}}\,
    {\rm e}^{-[x-\alpha(t)]^2/2\sigma^2(t)}\,,\qquad(t\geq t_0)
                       \eqno{(5.8)}$$
where we used the notation
$$\alpha(t)=x_0\e^{-\omega(t-t_0)}\,,\qquad
   \sigma^2(t)=\sigma_0^2\bigl[1-\e^{-2\omega(t-t_0)}\bigr]\,,
              \qquad(t\geq t_0)\,.\eqno{(5.9)}$$
The stationary Markov process associated to the
transition pdf (5.8) is selected by the initial, invariant pdf
$$f(x)={1\over\sigma_0\sqrt{2\pi}}\,\e^{-x^2/2\sigma_0^2}\eqno{(5.10)}$$
which is also the asymptotic pdf for every other initial condition when the
evolution is ruled by (5.7) (see [1]) so that the 
invariant distribution plays also the role of the limit distribution.
Since this invariant pdf also
coincides with the quantum stationary pdf $\phi_0^2=|\psi_0|^2$
the process associated by the SM to the ground state
of a quantum HO is nothing but the stationary Ornstein-Uhlenbeck process.
\par
For $n\geq1$ the solutions of (2.3) are no more so easy
to find and, as discussed in the previus section, we will have to solve the
eigenvalue problem (4.13) which, with
$\epsilon=\hbar\lambda$, can be written as
$$-\,{\hbar^2\over2m}\,G''(x)+
        \left({m\over2}\omega^2x^2-\hbar\omega\,{2n+1\over2}\right)G(x)
                    =\epsilon G(x)\,,\eqno{(5.11)}$$
in every interval $[x_k,x_{k+1}]$, with
$k=0,1,\dots,n$, between two subsequent singularities of the $v_n$ field.
The boundary conditions at the endpoints of these intervals, deduced
from (4.7) through (4.9), are
$$[\phi_nG'-\phi_n'G]_{x_k,x_{k+1}}=0\eqno{(5.12)}$$
and since $\phi_n$ (but not $\phi'_n$) vanishes in $x_k,x_{k+1}$, the
conditions to impose are
$$G(x_k)=G(x_{k+1})=0\eqno{(5.13)}$$
where it is understood that for $x_0$ and $x_{n+1}$ we
respectively mean
$$\lim_{x\to-\infty}G(x)=0\,,\qquad\lim_{x\to+\infty}G(x)=0\,.\eqno{(5.14)}$$
It is also useful at this point to give the eigenvalue problem in an
adimensional form by using the new adimensional variable $x/\sigma_0$
(which will still be called $x$) and the eigenvalue $\mu=\lambda/\omega=
\epsilon/\hbar\omega$. In this way the equation (5.11) with the conditions
(5.13) becomes
$$\eqalign{y''(x)-\left({x^2\over4}-{2n+1\over2}-\mu\right)y(x)&=0\cr
           y(x_k)=y(x_{k+1})&=0\cr}\eqno{(5.15)}$$
where $x, x_k, x_{k+1}$ are now adimensional variables. If $\mu_m$ and
$y_m(x)$ are the eigenvalues and eigenfunctions of (5.15), 
the general solution of the corresponding
FP equation (2.3) will be
$$f(x,t)=\sum_{m=0}^{\infty}c_m\e^{-\mu_m\omega t}
                  \phi_n(x)y_m\left({x\over\sigma_0}\right)\,.\eqno{(5.16)}$$
Of course the values of the coefficients $c_m$ will be fixed by the initial
conditions and by the obvious requirements that $f(x,t)$ must be non negative
and normalized (on the whole $x$ axis) along all its evolution.
Two linearly independent solutions of (5.15) are
$$y^{(1)}=\e^{-x^2/4}M\left(-\,{\mu+n\over2},{1\over2};{x^2\over2}\right)\,,
       \qquad y^{(2)}=
x\e^{-x^2/4}M\left(-\,{\mu+n-1\over2},{3\over2};{x^2\over2}\right)\,,
                                 \eqno{(5.17)}$$
where $M(a,b;z)$ are the confluent hypergeometric functions.
\par
We consider first the case $n=1$ ($x_0=-\infty$, $x_1=0$ and
$x_2=+\infty$) so that (5.15) will have to be solved separately for $x\leq0$
and for $x\geq0$ with the boundary conditions $y(0)=0$ and
$$\lim_{x\to-\infty}y(x)=\lim_{x\to+\infty}y(x)=0\,.\eqno{(5.18)}$$
A long calculation [5] shows that the transition pdf is now
$$p(x,t|x_0,t_0)={x\over\alpha(t)}\,
{\e^{-[x-\alpha(t)]^2/2\sigma^2(t)}-
        \e^{-[x+\alpha(t)]^2/2\sigma^2(t)}
           \over \sigma(t)\sqrt{2\pi}}\eqno{(5.19)}$$
where $\alpha(t)$ and $\sigma^2(t)$ are
defined in (5.9). It must be remarked however that (5.19)
must be considered as restricted to $x\geq0$ when
$x_0>0$ and to $x\leq0$ when $x_0<0$, and that only on these intervals
it is suitably normalized.
In order to take into account at once both these possibilities we can also 
introduce the Heavyside function $\Theta(x)$ so that for every $x_0\not=0$ we 
will have
$$p(x,t|x_0,t_0)=\Theta(xx_0)\,{x\over\alpha(t)}\,
{\e^{-[x-\alpha(t)]^2/2\sigma^2(t)}-
        \e^{-[x+\alpha(t)]^2/2\sigma^2(t)}
           \over \sigma(t)\sqrt{2\pi}}\,.\eqno{(5.20)}$$
This completely solves the problem for $n=1$ since from (4.8) we can now 
deduce also the evolution of every other initial pdf. In particular it can be 
shown that 
$$\lim_{t\to+\infty}p(x,t|x_0,t_0)=2\Theta(xx_0)\,
                {x^2\e^{-x^2/2\sigma_0^2}\over\sigma_0^3\sqrt{2\pi}}=
                        2\Theta(xx_0)\phi_1^2(x)\,,\eqno{(5.21)}$$
and hence, if $f(x,t_0^+)=f_0(x)$ is the initial pdf, we have for $t>t_0$
$$\eqalign{\lim_{t\to+\infty}f(x,t)&=\lim_{t\to+\infty}
                \int_{-\infty}^{+\infty}p(x,t|y,t_0)f_0(y)\,dy\cr
&=2\phi_1^2(x)\int_{-\infty}^{+\infty}\Theta(xy)f_0(y)\,dy
                =\Gamma(q;x)\phi_1^2(x)\,,\cr}\eqno{(5.22)}$$
where we have defined the function
$$\Gamma(q;x)=q\Theta(x)+(2-q)\Theta(-x)\,;\qquad 
                             q=2\int_0^{+\infty}f_0(y)\,dy\,.\eqno{(5.23)}$$
Remark that when $q=1$ (namely when the initial probability is equally shared 
on the two real semi-axis) we have $\Gamma(1;x)=1$ and the asymptotical pdf 
coincides with the quantum stationary pdf $\phi_1^2(x)$; if on the other hand 
$q\not=1$ the asymptotical pdf has the same shape of $\phi_1^2(x)$ but with 
different weights on the two semi-axis.
\par
If then $n=2$ we have $x_0=-\infty$, $x_1=-1$, $x_2=1$ and $x_3=+\infty$, 
and the equation (5.15) must be solved in the three intervals $(-\infty,-1]$,
$[-1,1]$ and $[1,+\infty)$, but the eigenvalues and eigenfunctions 
are now not easy to find so 
that a complete analysis of this case (and of every other case with
$n>2$) has still to be elaborated. At present only a few 
indications can be obtained numerically [5]:
for example it can be shown that, beyond $\mu_0=0$, 
the first eigenvalues in the interval $[-1,1]$ 
can be calculated as the first values such that
$$M\left(-\,{\mu+1\over2},{3\over2};{1\over2}\right)=0\eqno{(5.24)}$$
and are $\mu_1\sim 7.44$, $\mu_2\sim 37.06$, $\mu_3\sim 86.41$.
Also for the unbounded interval $[1,+\infty)$ (the analysis is similar for
$(-\infty,-1]$) the eigenvalues are derivable only numerically.
\vskip 30pt
\centerline{\bf 6. Controlled evolutions}
\vskip 10pt\noindent
It is important to remark now that solutions of the type (5.8) and (5.20),
and any other solution different from $|\phi_n|^2$, are
not associated to quantum mechanical states solutions of (2.10); in
other words, they define processes that satisfy neither the stochastic
variational principle [3] nor the Nelson dynamical equation [2].
That notwithstanding these processes still keep
an interesting relation with the quantum mechanics. In fact to every 
solution $f(x,t)$ of a FP equation (3.1), with a
given $v(x,t)$ and the constant diffusion coefficient (2.11), we can always 
associate the wave function of a quantum system if we take a suitable
time-dependent potential. This means in practice that even the
{\it virtual} (non optimal) processes discussed in this paper can be
associated to proper quantum states, namely can be made optimal
provided that the potential $V(x)$ of (2.10) be modified in a new $V(x,t)$ 
in order to control the evolution.
\par
Let us take a solution $f(x,t)$ of the FP equation (3.1), with
a given $v(x,t)$ and a constant diffusion 
coefficient (3.3): if we define the functions $R(x,t)$ and $W(x,t)$ from
$$f(x,t)=R^2(x,t)\,,\qquad\quad v(x,t)=\partial_x W(x,t)\,,\eqno{(6.1)}$$
if we remember from (2.12) that the following relation must hold
$$mv=\partial_x S+\hbar\,{\partial_x R\over R}=
            \partial_x S+{\hbar\over2}\,{\partial_x f\over f}=
        \partial_x\left(S+{\hbar\over2}\ln\tilde f \right)\eqno{(6.2)}$$
where $\tilde f$ is an adimensional pdf (it is the argument of a logarithm) 
obtained by means of a suitable and arbitrary multiplicative constant, and if
$S(x,t)$ is supposed to be the phase of a wave function as in (2.8),
we immediately get the equation
$$S(x,t)=mW(x,t)-{\hbar\over2}\ln \tilde f(x,t)-\theta(t)\eqno{(6.3)}$$
which allows us to determine $S$ from $f$ and $v$ (namely $W$) up to
an additive arbitrary function of the time $\theta(t)$. However, in order that
the wave function (2.8) with the said $R$ and $S$ be a solution of a
Schr\"odinger equation, we must also be sure that the
HJM equation (2.9) is satisfied. Since $S$ and $R$ are
now fixed, the equation (2.9) must be considered as a relation defining
a potential which, after a short calculation, becomes
$$V(x,t)={\hbar^2\over4m}\,\partial_x^2\ln \tilde f+
    {\hbar\over2}\left(\partial_t\ln \tilde f+v\partial_x\ln \tilde f\right)
      -{mv^2\over2}-m\partial_tW+\dot\theta\,.\eqno{(6.4)}$$
Of course if we start with a quantum wave function for a given potential
and if we pick up as a solution of (2.3) exactly $f=R^2$, the formula (6.4)
will correctly give back the initial potential, as can be seen for both the
ground state and the first excited state of the HO which
(by choosing respectively $\theta(t)=\hbar\omega t/2$ and
$\theta(t)=3\hbar\omega t/2$, which amounts to suitably fix the
zero of the potential energy) give as result the usual harmonic potential
(5.1).
\par
If on the other hand we consider for example 
the (non stationary) fundamental
solution (5.8) associated to the velocity field $v_0(x)$ of (5.6) for
the case $n=0$ of the HO (we put $t_0=0$ to simplify
the notation) we have already remarked that it does not correspond to
a quantum wave function whatsoever. However 
a short calculation shows that, by choosing
$$\dot\theta(t)=
    {\hbar\omega\over2}\left({2\sigma_0^2\over\sigma^2(t)}-1\right)=
         {\hbar\omega\over2}\,{1\over\tanh \omega t}\to
              {\hbar\omega\over2}\,,\qquad(t\to+\infty)\,,\eqno{(6.5)}$$
and the time-dependent controlling potential
$$V(x,t)={\hbar\omega\over2}\left[{x-\alpha(t)\over\sigma(t)}\right]^2
            {\sigma_0^2\over\sigma^2(t)}-{m\omega^2x^2\over2}\to
                {m\omega^2x^2\over2}\,,\qquad(t\to+\infty)\eqno{(6.6)}$$
we can define a quantum state (a wave function solution 
of a Schr\"odinger equation) which realizes the required evolution (5.8).
Of course the fact that for $t\to+\infty$ we recover the harmonic potential
is associated to the fact, aleady remarked, that the usual quantum pdf
$\phi_0^2(x)$ is also the limit distribution for every initial condition
and in particular also for the pdf (5.8).
In the case $n=1$, with $v_1(x)$ from (5.6) and the transition
probability (5.20) as given non-stationary solution, the calculations are
lenghtier. However if we define
$$\displaylines{\hfill F(x,t)={\e^{-[x-\alpha(t)]^2/2\sigma^2(t)}
                      \over\sigma(t)\sqrt{2\pi}}\,,\qquad
                G(x,t)={\e^{-[x+\alpha(t)]^2/2\sigma^2(t)}
                      \over\sigma(t)\sqrt{2\pi}}\,,\hfill(6.7)\cr
            \hfill T\left[{x\alpha(t)\over\sigma^2(t)}\right]=
{x\alpha(t)\over\sigma^2(t)}\,{F(x,t)+G(x,t)\over F(x,t)-G(x,t)}\,,\qquad\quad
          T(x)={x\over\tanh x}\,,\hfill(6.8)\cr}$$
and if we choose
$$\dot\theta(t)={\hbar\omega\over2}
        \left({4\sigma_0^2\over\sigma^2(t)}-
                  {2\sigma_0^2\alpha^2(t)\over\sigma^4(t)}-1\right)
         \to{3\over2}\,\hbar\omega\,,\qquad(t\to+\infty)\eqno{(6.9)}$$
we have as time dependent potential for every $x\not=0$
$$\eqalign{V(x,t)&={m\omega^2x^2\over2}
                \left({2\sigma_0^4\over\sigma^4}-1\right)+
                    \hbar\omega\left[1-{\sigma_0^2\over\sigma^2}\,
                       T\left({x\alpha\over\sigma^2}\right)\right]-
{\hbar^2\over4mx^2}\,\left[1-T\left({x\alpha\over\sigma^2}\right)\right]\cr
   &\to{m\omega^2x^2\over2}\,,\qquad\qquad(t\to+\infty)\,.\cr}\eqno{(6.10)}$$
In this case the asymptotic potential is the usual harmonic potential, but we 
must consider it separately on the positive and negative $x$ semi-axis since in 
the point $x=0$ a singular behaviour would show up
when $t\to0$. This means that, also if
asymptotically we recover the right potential, this will be 
associated with new 
boundary conditions in $x=0$ since we will be obliged to keep the system 
bounded on the positive (for example) semi-axis.
\vskip 30pt
\centerline{\bf 7. Modelling transitions}
\vskip 10pt\noindent
The explicit knowledge of the transition pdf of the type (5.8) and (5.20),
And the possibility of turning optimal any suitable $(f,v)$ state by
a right choice of $V(x,t)$
enable us also to explore the possibility of modelling evolutions leading,
for example, from the pdf of a given stationary state to another (decays and
excitations). In fact a spontaneuous generalization of this idea hints to
the possibility of modelling evolutions ffrom a given, arbitrary pdf and the pdf
of an eigenfuncition of some observable: something which could become
an element for 
very simple models of quantum measurements where we try to dynamically
describe the wave packet collapse.
As a first example let us consider the transition between the invariant pdf's
$$\eqalign{f_0(x)&=\phi_0^2(x) =
           {1\over\sigma_0\sqrt{2\pi}}\,\e^{-x^2/2\sigma_0^2}\,,\cr
f_1(x)&=\phi_1^2(x)=
           {x^2\over\sigma_0^3\sqrt{2\pi}}\,\e^{-x^2/2\sigma_0^2}\,.
                                                        \cr}\eqno{(7.1)}$$
If for instance we choose to describe the decay $1\to0$ we should just
use the Chapman-Kolmogorov equation (4.8) with (5.8) as transition pdf and
$f_1(x)$ as initial pdf ($t_0=0$). 
An elementary integration will show in this case
that the resulting evolution takes the form
$$f_{1\to0}(x,t)=\beta^2(t)f_0(x)+\gamma^2(t)f_1(x)\eqno{(7.2)}$$
where we used the notation
$$\beta^2(t)=1-\e^{-2\omega t}\,,\qquad 
                              \gamma(t)=\e^{-\omega t}\,.\eqno{(7.3)}$$
Taking now $v_0(x)$ from (5.6) and the evolving pdf from (7.2) 
and putting them in (6.4) (remark that, since $v_0$ is stationary, 
$\partial_t W=0$) we get the following form of the controlling potential:
$$V(x,t)={m\omega^2 x^2\over2}-2\hbar\omega U(x/\sigma_0;\beta/\gamma)
                                               \eqno{(7.4)}$$
where
$$U(x;b)={x^4+b^2x^2-b^2\over (b^2+x^2)^2}\,.\eqno{(7.5)}$$
In our example the parameter
$$b^2(t)={\beta^2(t)\over\gamma^2(t)}=\e^{2\omega t}-1\eqno{(7.6)}$$
is such that $b^2(0^+)=0$ and $b^2(+\infty)=+\infty$ and hence
$U$ goes everywhere to zero for $t\to+\infty$, but is everywhere 1 
with a negative singularity in $x=0$ for $t\to0^+$. As a consequence,
while for $t\to+\infty$ the controlling potential (7.4) behaves 
like the HO potential (5.1), for $t\to0^+$ it presents an unessential shift 
of $-2\hbar\omega$ in the zero level, but shows also 
a deep negative singularity in $x=0$.
\par
Apart from this singular behaviour of the controlling potential, a problem
arises from the form of the phase funcion $S$. 
In fact from (6.3) we easily have for our decay
$$S(x,t)=-{\hbar\over2}\ln\left[\beta^2(x,t)+{x^2\over\sigma_0^2}
                               \gamma^2(x,t)\right]
                           -{\hbar\omega\over2}\,t\eqno{(7.7)}$$
so that in particular we have
$$S(x,0^+)=-{\hbar\over2}\,\ln{x^2\over\sigma_0^2}\,,\eqno{(7.8)}$$
while we would have expected that initially our phase function 
be independent from $x$
as for every stationary wave function: this means that in our supposed evolution
the phase function presents a discontinuous behaviour for $t\to0^+$.
The problem arises here from the fact that in our simple model we initially
have a stationary state characterized by a ddp 
$f_1(x)$ and a velocity field $v_1(x)$,
and then suddenly, in order to start the decay, we suppose the same $f_1$
embedded in a different velocity field $v_0(x)$ which drags it toward a new
stationary $f_0(x)$. This discontinuous change from $v_1$ to $v_0$ is of course
responsible for the remarked discontinuous change in the 
phase of the wave function.
Hence a more realistic model for a controlled transition must take into account
a continuous and smooth (albeit widely arbitrary)
modification of the initial velocity field into the final one,
a requirement which compels us to consider a new class of FP equations
with time-dependent velocity field $v(x,t)$. 
In particular to achieve the proposed controlled
decay between two stationary states we should solve an evolution equation with a
velocity field $v(x,t)$ continuously, and possibly smoothingly, going 
from $v_1(x)$
to $v_0(x)$; but this seems at present beyond the reach of 
our possibilities since every
reasonable such $v(x,t)$ field has proven intractable from the 
point of view of the solution of 
the FP equation (2.3). However we can show the results for another meaningful
example which does not present the same technical difficulties of 
the decay between
two stationary states: namely the controlled evolution 
from a coherent oscillating 
packet in a HO, and the ground state of the same HO.
\par
To do this we will recall a simple result [1] which indicates how to find the 
solutions of a particular class of evolution 
equations (2.3) which contains the situation of
our proposed example. 
If the velocity field of the evolution equation (2.3) has the linear form
$$v(x,t)=A(t)+B(t)x\eqno{(7.9)}$$
with $A(t)$ and $B(t)$ continuous functions of time,
then there are always solutions of the form
${\cal N}\bigl(\mu(t),\nu(t)\bigr)$ where $\mu(t)$ and $\nu(t)$ are
calculated from the differential equations
$$\mu'(t)-B(t)\mu(t)=A(t)\,;\qquad
\nu'(t)-2B(t)\nu(t)=2D\eqno{(7.10)}$$
with suitable initial conditions.
On the other hand the (non stationary) wave 
function of the oscillating coherent wave
packet with initial displacement $a$ is
$$\psi_c(x,t)=\biggl({1\over2\pi\sigma_0^2}\biggr)^{1/4}
\exp\biggl[-{(x-a\cos\omega t)^2\over4\sigma_0^2}
-i\biggl({4ax\sin\omega t-a^2\sin2\omega t\over8\sigma_0^2}+{\omega t\over2} 
                                           \biggr)\biggr] \eqno{(7.11)}$$
so that the corresponding forward velocity field will be
$$v_c(x,t)=a\omega(\cos\omega t-\sin\omega t)-\omega x\,,\eqno{(7.12)}$$
namely it will have the required form (7.9) 
with $A(t)= a\omega(\cos\omega t-\sin\omega t)$
and $B(t)=-\omega$, while the position pdf will be
$$f_c(x,t)=|\psi_c(x,t)|^2=f_0(x-a\cos\omega t)\,. \eqno{(7.13)}$$
Now it is very easy to show that when $B(t)=-\omega$, as
in the case of our wave packet, there are stable,
coherent (non dispersive) solutions with $\nu(t)=\sigma_0^2$ of the form
${\cal N}\bigl(\mu(t),\sigma_0^2\bigr)$, namely of the form
$$f(x,t)=f_0\bigl(x-\mu(t)\bigr)\,. \eqno{(7.14)}$$
Of course the time evolution of such coherent solutions can be determined
in one step, without implementing the two steps procedure of first calculating
the transition pdf and then, through the 
Chapman-Kolmogorov equation, the evolution
of an arbitrary initial pdf.
On the other hand if we compare (5.6) and (7.12) we see 
that the difference between
$v_0$ and $v_c$ consists in the first, time dependent 
term of the second one; hence
it is natural to consider the problem of solving the evolution equation (2.3) 
with a velocity field of the type 
$$\eqalign{v(x,t)&=A(t)-\omega x\cr
           A(t)&=a\omega(\cos\omega t-\sin\omega t)F(t)\cr}\eqno{(7.15)}$$
where $F(t)$ is an arbitrary function varying smoothly between 1 and 0, 
or vice verssa.
In this case the evolution equation (2.3)
still has stable, coherent (non dispersive) solutions of the form (7.14)
with a $\mu(t)$ dependent on our choice of $F(t)$ through (7.10).
\par
A completely smooth transition from the coherent, oscillating wave function 
(7.11) to the ground state $\phi_0$ (5.4) of the HO can now be achieved 
for example by means of the following choice of the function $F(t)$:
$$F(t)=1-\left(1-\e^{-\Omega t}\right)^N
           =\sum_{k=1}^N(-1)^{k+1}{N\choose k}\e^{-\omega_k t}\eqno{(7.16)}$$
where 
$$\Omega={\ln N\over \tau}\,,\qquad\omega_k=k\Omega\,;
                            \qquad \tau>0\,,\qquad N\geq 2\,.\eqno{(7.17)}$$
In fact this $F(t)$ goes monotonically from $F(0)=1$ to $F(+\infty)=0$ with a
flex point in $\tau$ (which can be considered as 
the arbitrary instant of the transition)
where its derivative $F'(\tau)$ is negative and grows, in absolute value, 
logarithmically with $N$. The condition $N\geq 2$ 
also guarantees that $F'(0)=0$,
and hence that the controlling potential $V(x,t)$ of (6.4) 
will continuously start at
$t=0$ from the HO potential (5.1), and eventually come back to it for
$t\to+\infty$. Finally the phase function $S(x,t)$ 
too will change continuously from
that of $\psi_c$ to that of the HO ground state. A long but simple calculation
will now show that the explicit form of the controlling potential is
$$V(x,t)=m\omega^2{x^2\over2}-m\omega ax\sum_{k=1}^N(-1)^{k+1}{N\choose k}
   \big[U_k(t)\omega_k\e^{-\omega_k t}- W_k\omega\e^{-\omega t}\big] 
                             \eqno{(7.18)}$$
where
$$\eqalign{U_k(t)&=\sin\omega t+{2\omega^2\sin\omega t-\omega_k^2\cos\omega t
                      \over (\omega_k-\omega)^2+\omega^2}\,,\cr
                  W_k&=1+{2\omega^2-
         \omega_k^2\over(\omega_k-\omega)^2+\omega^2}
              =\sqrt{2}U_k\left({\pi\over4\omega}\right)\,.
                                        \cr} \eqno{(7.19)}$$
The parameters $\tau$ and $N$, with the limitations (7.17), are free 
and connected to the
particular form of the transition that we want to implement.
We conclude this section by remarking that, in a HO, the 
transition between a coherent, 
oscillating wave packet and the ground state is a 
transition between a (Poisson) 
superposition of all the energy eigenstates to just one 
energy eigenstate: an outcome
which is similar to that of an energy measurement, but for 
the important fact that here the
result (the energy eigenstate) is deterministically 
controlled by a time dependent potential.
In fact our controlled transition does not produce 
mixtures, but pure states (eigenstates) and
in some way realizes a dynamical model for one
of the branches of a measurement leading to an eigenvalue and an eigenstate.
\vskip 30pt
\centerline{\bf 8. Beam dynamics in particle accelerators}
\vskip 10pt\noindent
As a model which tries to put in evidence the classical aspects
of the quantum physics, the SM seems especially suitable to
the description of systems whose nature in some sense lies between
classical and quantum: the so called mesoscopic or quantum-like
systems [15].
We will propose now a few preliminary remarks about the
possibility of making use of this characteristic
in a particular physical domain [7].
The dynamical evolution of beams in particles accelerators 
is a typical example of mesoscopic behaviour. Since they are governed
by external electromagnetic forces and by the interaction of the beam
particles among themselves and with the environment, 
charged beams are higly nonlinear dynamical
systems, and most of the studies on colliding beams 
rely either on classical phenomena such as nonlinear resonances, 
or on isolated sources of unstable behaviors
as building blocks of more complicated chaotic instabilities.
This line of inquiry has produced a general qualitative
picture of dynamical processes in particle accelerators
at the classical level.
However, the coherent oscillations of the beam density and 
profile require, to be explained, some mechanism of
local correlation and loss of statistical independence.
This fundamental observation points towards the need 
to take into account all the interactions as a whole. 
Moreover, the overall interactions
between charged particles and machine elements
are really nonclassical in the sense that of the
many sources of noise that are present, almost all
are mediated by fundamental quantum processes of
emission and absorbtion of photons.
Therefore the equations describing these processes 
must be, in principle, quantum.  
\par
Starting from the above considerations,
two different approaches to the classical collective 
dynamics of charged beams have been developed, one
relying on the FP equation [16]
for the beam density, another based on a mathematical
coarse graining of Vlasov equation leading to a
quantum-like Schr\"odinger equation, with a thermal
unit of emittance playing the role of Planck constant [17].
The study of statistical effects on the dynamics
of electron (positron) colliding beams
by the FP equation has led to several interesting
results, and has become an established reference in treating
the sources of noise and dissipation in particle accelerators
by standard classical probabilistic techniques [18].
Concerning the relevance of the quantum-like approach, 
at this stage we only want to point out that some 
recent experiments on confined classical systems subject
to particular phase-space boundary conditions seem to
to be well explained by a quantum-like 
(Schr\"odinger equation) formalism [19].
In this approach [20]
the (one dimensional)
transverse density profile of the beam
is described in terms of a complex function, called beam wave function, 
whose squared modulus give the transverse density profile of the beam.
This beam wave function satisfies a Schr\"odinger-like equation where
$\hbar$ is replaced by the transverse beam emittance $\epsilon$:
$$i\epsilon{\partial \psi(x,z)\over\partial z}=-{\epsilon^2\over2}
   {\partial^2\psi(x,z)\over\partial x^2}+U(x,z)\psi(x,z)\,.\eqno{(8.1)}$$
On the other hand a recently proposed model for the description
of collective beam dynamics in the semiclassical regime [21] 
relies on the idea of
simulating semiclassical corrections to
classical dynamics by suitable classical stochastic 
fluctuations with long range coherent correlations,
whose scale is ruled by Planck constant.
This elaborates a hypothesis first proposed by Calogero [22]
in his attempt to prove that quantum mechanics might
be interpreted as a tiny chaotic component of 
the individual particles' motion in a gravitationally
interacting universe.
The virtue of the proposed semiclassical model is twofold:
on the one hand it can be formulated both in a probabilistic 
FP fashion and in a quantum-like 
(Schr\"odinger) setting, thus bridging the formal
gap between the two approaches. On the other hand
it goes further by
describing collective effects beyond the classical regime due 
to the semiclassical quantum corrections. 
\par
Since we are interested in the description of the stability
regime, when thermal dissipative effects are
balanced on average by the RF energy pumping, and the
overall dynamics is conservative and time-reversal
invariant in the mean, the choice to model the random
kinematics with 
the Nelson diffusions, that are nondissipative and 
time-reversal invariant, is particularly natural.
The diffusion process describes the effective motion
at the mesoscopic level (interplay of thermal equilibrium,
classical mechanical stability, and fundamental quantum 
noise) and therefore the diffusion coefficient is set to be 
the semiclassical unit of emittance provided by qualitative 
dimensional analysis. In other words, we simulate the quantum
corrections to classical deterministic motion (at leading order
in Planck constant) with a suitably defined random kinematics
replacing the classical deterministic trajectories.
Therefore, apart from the different objects
involved (beam spatial density versus Born probability density;
Planck constant versus emittance), the dynamical equations of 
our model formally reproduce the equations of the Madelung
fluid (hydrodynamic) representation of quantum mechanics. 
In this sense, the present scheme allows for a 
quantum-like formulation equivalent to the probabilistic one.
\par
With a few changes in the notation we can now reproduce, for
the beam dynamics, the SM approach sketched in section 2. Let 
$q(t)$ be the process representing some collective degree of 
freedom of the beam with a pdf $\rho(x,t)$. 
Then, in suitable units, the basic 
stochastic kinematical
relation is an It\^o stochastic differential equation
of the type (2.1) where the emittance $\epsilon$ of the beam
plays the role of a diffusion coefficient.
Since we are interested in the stability regime of the bunch
oscillations, the bunch itself can be 
considered in a quasi-stationary state, during which the energy lost 
by dissipation is regained in the RF cavities. In such a quasi-stationary
regime the dynamics is,
on average, invariant for time-reversal
and we can define a classical effective Lagrangian 
$L(q, {\dot{q}})$ of the system, where the 
classical deterministic kinematics is replaced by the random
diffusive kinematics (2.1). 
The equations for the dynamics can then be obtained from the classical
Lagrangian by means of the stochastic variational principles. 
\par 
Introducing now the time-like coordinate $s=ct$ we get now the
analog of the equations (2.3) and (2.6) in the form of a 
HJM equation
$$\partial_sS+{v^2\over2}-2\epsilon^2
        {\partial^2_x\sqrt{\rho}\over\sqrt{\rho}}+V(x,s)=0\,,
                      \eqno{(8.2)}$$
and of a continuity equation
$$\partial_s\rho=-\partial_x(\rho v)\,.\eqno{(8.3)}$$
Remark that now the symbol $v$ no more represents the forward
velocity fields, but rather the drift velocity connected to the
forward and backward velocities by the relation
$2v=v_{(+)}+v_{(-)}$, and to the phase function by the relation
$v=\partial_xS$.
The observable structure is now quite clear:
$\E v$ is the average velocity of 
the bunch center oscillating along the transverse direction;
$\E q$ gives the average coordinate of the bunch center;
finally the second moment $(\Delta q)^2 = E\big(q-E(q)\big)^2$
determines the dispersion
(spreading) of the bunch. The
coupled equations of dynamics may now be used to achieve a controlled
coherence: given a desired state $(\rho, v)$ the equations of motion
(8.2) and (8.3) can be solved to calculate the external controlling potential 
$V(x,s)$ that realizes this state.
\par
General techniques to obtain localized quantum wavepackets as dynamically
controlled systems in SM have already been introduced [23].
In this way one can construct for general systems either coherent
packets following the classical trajectories with constant dispersion, or
coherent packets following the classical trajectories with time-dependent,
but at any time bounded dispersion. These results can now be extended also 
to the quantum-like description of the transverse dynamics of a particle beam
and hence it will be possible to select a current velocity, by fixing the
characteristics of the motion of the packet center, to determine the 
corresponding solutions of the FP (continuity) equation and
finally to use the HJM equation as a constraint giving
us the controlling device. The formal details of this program will be developed
in a subsequent paper.
\vskip 30pt
\centerline{\bf 9. Concluding remarks}
\vskip 10pt\noindent
It has been observed that the inverse problem of determining a 
controlling potential for a given quantum evolution in fact does not need
to be formulated in terms of SM. Given two quantum wave function $\psi_1$
and $\psi_2$ we could indeed design a new wave function $\psi(x,t)$,
evolving from $\psi_1$ to $\psi_2$ plug it, as required evolution,
directly in the Schr\"odinger equation (2.10) and eventually deduce from
that the form of the controlling potential. At first glance this seems 
to completely circumwent the need for a model like the SM: given an
arbitrary evolving state we can always calculate the potential producing
it. However about that two remarks are in order.
\par
First of all, from a purely technical point of view, the simplification
introduced by this procedure shows up to be elusive. In fact we must
remember that a quantum wave function has complex values and hence, if
we simply take an arbitrary evolution, the resulting potential calculated
from the Schr\"odinger equation (2.10) will also be complex. This means that,
to have a real valued potential, we must impose some conditions on the
supposed evolution. These conditions of course depend on the hypothesized
form of $\psi$. For example, if we fix the evolution of its modulus, the
said condition will materialize in a partial differential equation on
the phase function $S$ of the wave function. On the other hand the use
of the HJM  equation (2.9) as the tool to solve the 
inverse problem always give a real valued potential as a result.
\par
However both the two proposed procedures are possible and, to identically
posed questions, they will give identical answers. Given this
obvius equivalence, the second remark is that our choice of the procedure
will be operated on the basis of opportunity considerations. 
In both cases the result will be influenced by the starting
hypothesis on the supposed evolution of the state $\psi$ modelling the
transition from $\psi_1$ to $\psi_2$. But, since the observable part of
the wave function is its square modulus, namely the position
pdf, the relevant hypothesis will
be on its evolution. The phase function, or, equivalently, the velocity 
fields, are not directly observable, and hence
are at first sight of secondary concern. Their importance
become apparent only when we require that the potential be real or
that the transitions show a realistic, smooth behaviour. Hence, depending
on the specific problem we are dealing with, it could be more suitable
to approach it in terms of a state given through a wave function
$\psi$, or in
terms of a state given through the couple $(f,v)$. The two approaches
are certainly equivalent, but one may prove to be more suggestive.
In particular that based on the SM equations seems to be better
for the treatment of systems, like as the mesoscopic, quantum-like ones,
which are well described by classical probabilistic models in terms 
of real space-time trajectories.
\vskip 30pt\noindent
{\bf REFERENCES}
\vskip 10pt
{\baselineskip 12pt
\item{1.}{N.Cufaro Petroni and F.Guerra: 
{\it Found.Phys.} {\bf 25} (1995) 297;}
\item{}  {N.Cufaro Petroni: Asymptotic behaviour of densities 
          for Nelson processes, in 
         {\it Quantum communications and measurement}, V.P.Belavkin et 
          al. Eds., 
          Plenum Press, New York, 1995, p. 43;}
\item{}  {N.Cufaro Petroni, S.De Martino and S.De Siena: Non 
          equilibrium densities of Nelson
          processes, in {\it New perspectives in the physics of 
          mesoscopic systems}, 
          S.De Martino et al. Eds., World Scientific, Singapore, 1997, p. 59.}
\item{2.}{E.Nelson: {\it Phys.Rev.} {\bf 150} (1966) 1079;}
\item{}  {E.Nelson: {\it Dynamical Theories of Brownian Motion} (Princeton
          U.P.; Princeton, 1967);}
\item{}  {E.Nelson: {\it Quantum Fluctuations} (Princeton U.P.;
          Princeton, 1985);}
\item{}  {F.Guerra: {\it Phys.Rep.} {\bf 77} (1981) 263.}
\item{3.}{F.Guerra and L.Morato: {\it Phys.Rev.} {\bf D 27} (1983) 1774;}
\item{}  {F.Guerra and R.Marra: {\it Phys.Rev.} {\bf D 28} (1983) 1916;}
\item{}  {F.Guerra and R.Marra: {\it Phys.Rev.} {\bf D 29} (1984) 1647.}
\item{4.}{D.Bohm and J.P.Vigier: {\it Phys.Rev.} {\bf 96} (1954) 208.}
\item{5.}{N.Cufaro Petroni, S.De Martino and S.De Siena: Exact solutions of
          Fokker-Planck equations associated to quantum wave functions; in
          press on Phys.Lett. {\bf A}.}
\item{6.}{N.Cufaro Petroni: {\it Phys.Lett.} {\bf A141} (1989) 370;}
\item{}  {N.Cufaro Petroni: {\it Phys.Lett.} {\bf A160} (1991) 107;}
\item{}  {N.Cufaro Petroni and J.P.Vigier: {Found.Phys.} {\bf 22} (1992) 1.}
\item{7.}{N.Cufaro Petroni, S.De Martino, S.De Siena and F.Illuminati:
          A stochastic model for the semiclassical collective dynamics of
          charged beams in partcle accelerators; contribution to the 15th ICFA 
          Advanced Beam Dynamics Workshop, Monterey (California, US), Jan 98.}
\item{}  {N.Cufaro Petroni, S.De Martino, S.De Siena, R. Fedele, F.Illuminati 
          and S. Tzenov: Sochastic control of beam dynamics; contribution to
          the EPAC'98 Conference, Stockholm (Sweden), Jun 98.} 
\item{8.}{E.Madelung: {\it Z.Physik} {\bf 40} (1926) 332;}
\item{}  {L.de Broglie: {\it C.R.Acad.Sci.Paris} {\bf 183} (1926) 447;}
\item{}  {L.de Broglie: {\it C.R.Acad.Sci.Paris} {\bf 184} (1927) 273;}
\item{}  {L.de Broglie: {\it C.R.Acad.Sci.Paris} {\bf 185} (1927) 380;}
\item{}  {D.Bohm: {\it Phys.Rev.} {\bf 85} (1952) 166, 180.}
\item{9.}{L.de la Pe\~na and A.M.Cetto: {\it Found.Phys.} {\bf 5} (1975) 355;}
\item{}  {N.Cufaro Petroni and J.P.Vigier: {\it Phys.Lett.} 
         {\bf A73} (1979) 289;}
\item{}  {N.Cufaro Petroni and J.P.Vigier: {\it Phys.Lett.}   
         {\bf A81} (1981) 12;}
\item{}  {N.Cufaro Petroni and J.P.Vigier: {\it Phys.Lett.} 
         {\bf A101} (1984) 4;}
\item{}  {N.Cufaro Petroni, C.Dewdney, P.Holland, T.Kyprianidis and J.P.Vigier: 
         {\it Phys.Lett.} {\bf A106} (1984) 368;}
\item{}  {N.Cufaro Petroni, C.Dewdney, P.Holland, T.Kyprianidis and J.P.Vigier: 
         {\it Phys.Rev.} {\bf D32} (1985) 1375.}
\item{10.}{F.Guerra: The problem of the physical interpretation of Nelson
          stochastic mechanics as a model for quantum mechanics, in {\it New 
          perspectives in the physics of mesoscopic systems}, 
          S.De Martino et al. Eds., World Scientific, Singapore, 1997, p. 133.}
\item{11.}{H.Risken: {\it The Fokker-Planck equation} (Springer, Berlin, 1989).}
\item{12.}{F.Tricomi: {\it Equazioni differenziali} (Einaudi, Torino, 1948).}
\item{13.}{F.Tricomi: {\it Integral equations} (Dover, New York, 1985).}
\item{14.}{S.Albeverio and R.H\o gh-Krohn: {\it J.Math.Phys.} {\bf 15} (1974)
          1745.}
\item{15.}{{\it New 
          perspectives in the physics of mesoscopic systems}, 
          S.De Martino et al. Eds., World Scientific, Singapore, 1997.}
\item{16.}{F. Ruggiero, {\it Ann.Phys.} (N.Y.) {\bf 153}, (1984) 122;}
\item{}   {J. F. Schonfeld, {\it Ann.Phys.} (N.Y.) {\bf 160}, (1985) 149.}
\item{17.}{R. Fedele, G. Miele and L.
          Palumbo, {\it Phys.Lett.} {\bf A194}, (1994) 113.}
\item{18.}{S. Chattopadhyay, AIP Conf. Proc. {\bf 127}, 444 (1983);}
\item{}   {F. Ruggiero, E. Picasso and L. A. 
          Radicati, {\it Ann.Phys.} (N. Y.) {\bf 197}, (1990) 396.}
\item{19.}{R. K. Varma, in {\it Quantum-like Models and Coherence Effects},
          R. Fedele et al. Eds. World Scientific, Singapore, 1996.}
\item{20.}{S.De Nicola, R.Fedele, G.Miele and V.Man'ko, in {\it New 
          perspectives in the physics of mesoscopic systems}, 
          S.De Martino et al. Eds., World Scientific, Singapore, 1997, p. 89.}
\item{21.}{S. De Martino, S. De Siena and F. Illuminati, {\it Mod.Phys.Lett.}
          {\bf B12} (1998), in press.}
\item{22.}{F. Calogero, {\it Phys.Lett.} {\bf A228}, (1997) 335.}
\item{23.}{S.De Martino, S. De Siena and F. Illuminati,
          {\it J.Phys.} {\bf A30} (1997) 4117.}
\par}
\bye